\documentclass[journal]{IEEEtran}
\usepackage{amsthm,amsmath,amssymb,mathtools,bm,etoolbox,float,hyperref}
\usepackage[dvipsnames]{xcolor}
\usepackage{graphicx,cite,cuted}
\usepackage{bigints}
\usepackage{caption}

\usepackage{stackengine}

\usepackage{mathrsfs,verbatim}
\usepackage{ellipsis}
\usepackage{tikz}
\usepackage{tikz-3dplot}
\usepackage[utf8]{inputenc}
\usepackage{pgfplots} 
\usepackage{pgfgantt}
\usepackage{pdflscape}
\pgfplotsset{compat=newest} 
\pgfplotsset{plot coordinates/math parser=false}
\pgfplotsset{every  tick/.style={black,},ylabel style={font=\tiny},xlabel style={font=\tiny},tick label style={font=\tiny},legend style= {font=\scriptsize},
minor x tick num=1,minor y tick num=1,xminorticks=true,yminorticks=true,}
\newlength\fheight
\newlength\fwidth
\usepackage{mathrsfs}

\usepackage{algorithm}
\usepackage{algpseudocode}
\makeatletter
\renewcommand{\Function}[2]{%
  \csname ALG@cmd@\ALG@L @Function\endcsname{#1}{#2}%
  \def\jayden@currentfunction{#1}%
}
\newcommand{\funclabel}[1]{%
  \@bsphack
  \protected@write\@auxout{}{%
    \string\newlabel{#1}{{\jayden@currentfunction}{\thepage}}%
  }%
  \@esphack
}

\usepackage{xinttools} 
\usepackage{tikz}
\usepackage{tkz-euclide}
\usetikzlibrary{calc}
\newtheorem{theorem}{Theorem}
\newtheorem{proposition}{Proposition}
\newtheorem{lemma}{Lemma}
\newtheorem{remark}{Remark}

\usepackage[dvipsnames]{xcolor}
\definecolor{cornellred}{rgb}{0.7, 0.11, 0.11}

\def\biglen{20cm} 
\tikzset{
  half plane/.style={ to path={
       ($(\tikztostart)!.5!(\tikztotarget)!#1!(\tikztotarget)!\biglen!90:(\tikztotarget)$)
    -- ($(\tikztostart)!.5!(\tikztotarget)!#1!(\tikztotarget)!\biglen!-90:(\tikztotarget)$)
    -- ([turn]0,2*\biglen) -- ([turn]0,2*\biglen) -- cycle}},
  half plane/.default={1pt}
}

\DeclareMathAlphabet{\pazocal}{OMS}{zplm}{m}{n}

\usepackage{caption}
\usepackage{multicol,subcaption}
\DeclareMathOperator*{\argmin}{arg\,min}

\begin{document}

\title{Full-Duplex Wideband mmWave Integrated Access and Backhaul with Low Resolution ADCs}



\author{Elyes~Balti,~\IEEEmembership{Student~Member,~IEEE,}
        and~Brian~L.~Evans,~\IEEEmembership{Fellow,~IEEE}
\thanks{Authors are with the Wireless Networking and Communications Group, Dept. of Electrical and Computer Eng., The University of Texas at Austin, Austin, TX 78712 USA (e-mails: ebalti@utexas.edu, bevans@ece.utexas.edu).}
}

\maketitle

\begin{abstract}
We consider a wideband integrated access and backhaul system operating in full-duplex mode between the New Radio gNB donor and single user equipment. Due to high power consumption in millimeter wave systems, we use low-resolution analog-to-digital converters (ADCs) in the receivers.  Our contributions include (1) hybrid beamformer to maximize sum spectral efficiency of the access and backhaul links by canceling self-interference and maximizing received power; (2) all-digital beamformer and upper bound on sum spectral efficiency; and (3) simulations to compare full vs. half duplex, finite vs. infinite ADC resolution, hybrid vs. all-digital beamforming, and the upper bound in spectral efficiency.
\end{abstract}


\section{Introduction}
\IEEEPARstart{F}{uture} wireless networks are expected to be highly dense to support the high standards of future applications, such as the Internet of Things, virtual/augmented reality, edge computing, vehicle-to-everything. However, traditional fiber backhauling is often an economically impractical solution for carrier operators. In this context, integrated access and backhaul (IAB) technology has emerged as a cost-effective alternative to the traditional fiber-backhauled system. In the case of IAB, only a few of the BSs are connected to the traditional wired infrastructures while the other BSs relay the
backhaul traffic wirelessly \cite{4,release17}. In a typical IAB framework, the access and backhaul links share the same frequency spectrum, which results in a resource collision problem; thus, resource management is required to resolve this issue. Owing to the simplicity of implementation, many previous studies have incorporated half duplex (HD) constraints in their frameworks \cite{4,release17}, which we refer to as HD IAB. In HD IAB, the access and backhaul links must use the given radio resources orthogonally, be it time or frequency. While this helps prevent collisions between the two separate links, it fails to exploit the full potential of the given radio resources 

In contrast, a smarter IAB framework with full duplex
(FD) techniques may simply rule out the HD constraint. FD  systems have recently gained enormous attention in academia and industry due to its potential to reduce latency and double spectral efficiency in the link budget compared to the half-duplex relays that transmit and receive in different time slots \cite{thesis}. These benefits make FD applicable in practice such as machine-to-machine and integrated access and backhaul which is currently proposed in 3GPP Release 17 \cite{release17}. 

Although FD brings many advantages, it suffers from loopback self-interference (SI), which is caused by the simultaneous transmission and reception over the same resource blocks. This loopback signal cannot be neglected as the relative SI power can be several orders of magnitude stronger than the signal power from the user equipment (UE), which can render FD systems dysfunctional \cite{zf}. 

In this letter, we consider a wideband FD IAB system with low-resolution ADCs in the receivers. To address the SI issues, we propose a robust hybrid beamforming design to cancel the SI and maximize the sum spectral efficiency. We also derive the all-digital solution and the upper bound and compare full vs. half duplex, finite vs. infinite ADC resolution, hybrid vs. all-digital beamforming and the upper bound.

The rest of the letter is organized as follows: Section II describes the system model including channels and signals under low-resolution ADCs. Section III presents the optimization problem as well as the beamforming design while Section IV discusses numerical results. Section V concludes the letter.

\textbf{Notation}: Bold lowercase $\mathbf{x}$ denotes column vectors, bold uppercase $\mathbf{X}$ denotes matrices, non-bold letters $x, X$ denote scalar values, and calligraphic letters $\mathcal{X}$ denote sets. Using this notation, $\|\mathbf{x}\|_2$ is the $\ell_2$ norm, $\| \mathbf{X}\|_F$ is the Frobenius norm, $\sigma_\ell(\mathbf{X})$ is the $\ell$-th singular value of $\mathbf{X}$ in decreasing order, $\mathbf{X}^*$ is the Hermitian or conjugate transpose, $\mathbf{X}^T$ is the matrix transpose, $\mathbf{X}^{-1}$ denotes the inverse of a square non-singular matrix and $[\mathbf{X}]_{mn}$ is the entry in the $m$-th row and $n$-th column of the matrix $\mathbf{X}$. We use $\mathbb{E}[\cdot]$ to denote the expectation.
\section{System Model}
\begin{figure}[t]
    \centering
    \includegraphics[width=\linewidth]{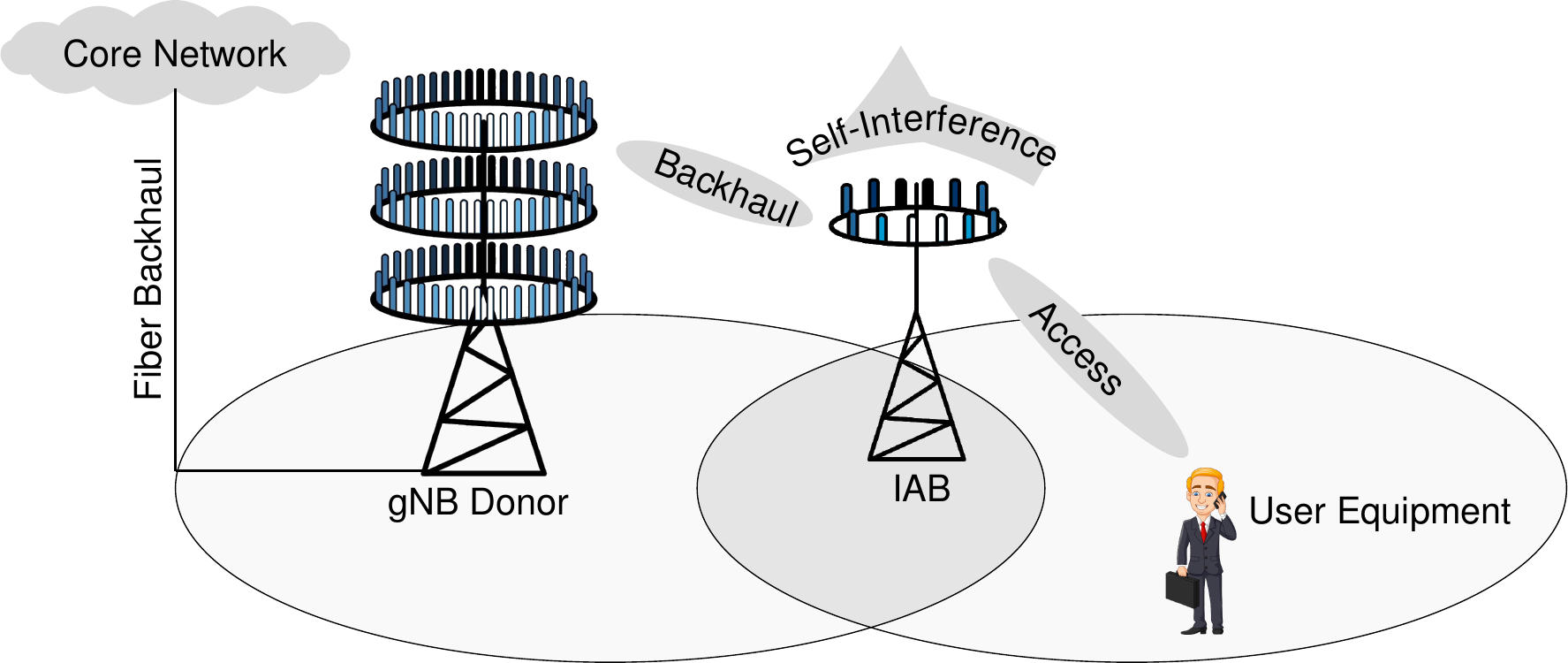}
    \caption{Illustration of full-duplex integrated access and backhaul (IAB) for a single-user case. The gNB donor, linked to the core network with fibered backhaul, communicates with the IAB node through wireless backhaul. The UE is served by the IAB node through the access link. Simultaneous transmission and reception of the IAB node over the same time/frequency resources blocks incurs the loopback self-interference signal. }
    \label{iab}
\end{figure}
Assuming that the maximum delay spread of the channel is within the cyclic prefix (CP) duration, we refer to OFDM waveform with $K$ subcarriers for wideband transmission. At the $k$-th subcarrier, the symbols $\mathbf{s}[k]$ are transformed into the time domain using the $K$-point Inverse Discrete Fourier Transform (IDFT). The CP of length ($L_c$) is then appended to the time-domain QAM symbols before applying the precoder. The OFDM block is formed by the CP followed by the $K$ time domain symbols with covariance matrix $\mathbb{E}[\mathbf{s}[k]\mathbf{s}^*[k]] = \frac{1}{KN_s}\mathbf{I}$, where $N_s$ is the number of allowable spatial streams.
\begin{remark}
The description of the OFDM wideband transmission is applicable to backhaul and access scenarios.
\end{remark}

\subsection{Channel Model}
In this work, we assume that the MIMO channels for backhaul and access are wideband, having a delay tap length $L$ in the time domain. The omnidirectional continuous time channel impulse response (CIR) can be expressed as
\begin{equation}
\mathsf{h}_{\mathsf{omni}}(t,\theta,\phi) =  \sum_{c=0}^{C-1}\sum_{r_c=0}^{R_c-1}\alpha_{r_c}\delta(t-\tau_{r_c})\delta(\theta-\theta_{r_c}) \delta(\phi-\phi_{r_c})    
\end{equation}
The continuous time CIR is not bandlimited.  When it is convolved with the pulse shaping filter impulse response $p(\tau)$, however, the CIR becomes bandlimited and hence can be sampled at rate $1/T_s$ to obtain the discrete-time channel. Equivalently, the $\ell$-th ($\ell = 0,\ldots,L-1$) tap delay of the discrete time baseband channel is given by
\begin{equation}\label{channel}
\begin{split}
\mathsf{H}[\ell] = \gamma \sum_{c=0}^{C-1}\sum_{r_c=0}^{R_c-1} \alpha_{r_c} p(\ell T_s -\tau_{r_c}) \mathbf{a}_{\mathsf{RX}}(\theta_{r_c})   \mathbf{a}_{\mathsf{TX}}^*(\phi_{r_c})    
\end{split}
\end{equation}
where $\gamma = \sqrt{\frac{N_{\mathsf{RX}}N_{\mathsf{TX}}}{CR_c}}$, $T_s$ is the signaling interval, $\theta_{r_c}$ and $\phi_{r_c}$ are the angle of arrival (AoA) and and the angle of departure (AoD) of the $r_c$-th ray, respectively. Each ray has a relative time delay $\tau_{r_c}$, and a complex path gain $\alpha_{r_c}$. Here $p(\tau)$ is a  raised cosine pulse evaluated at $\tau$. In addition, $\mathbf{a}_{\mathsf{RX}}(\theta)$ and $\mathbf{a}_{\mathsf{TX}}(\phi)$ are the antenna array response vectors of the RX and TX, respectively. The RX array response vector is given by
\begin{equation}
\mathbf{a}_{\mathsf{RX}}(\theta) = \frac{1}{\sqrt{N_{\mathsf{RX}}}}\left[1,\mathsf{e}^{j\frac{2\pi d}{\lambda} \sin(\theta)},\ldots,\mathsf{e}^{j\frac{2\pi d}{\lambda}\left(N_{\mathsf{RX}} -1\right)\sin(\theta)}  \right]^{T}.    
\end{equation}

The channel at the $k$-th subcarrier is given by
\begin{equation}
\mathbf{H}[k] = \sum_{\ell=0}^{L-1} \mathsf{H}[\ell] \mathsf{e}^{-j\frac{2\pi k}{K}\ell} 
\end{equation}
where $k=0,\ldots,K-1$.
\begin{figure*}[t]
    \centering
    \includegraphics[width=\linewidth]{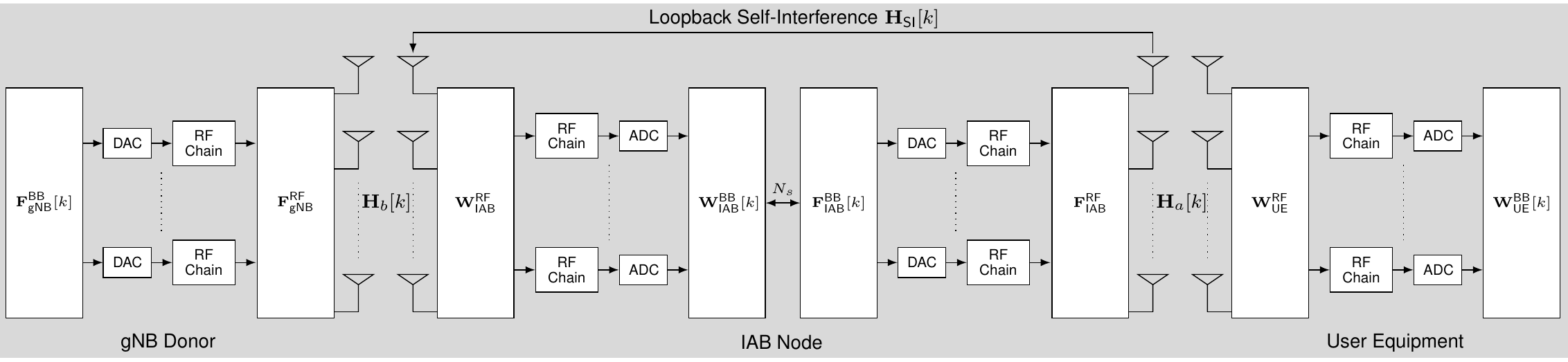}
    \caption{Basic abstraction of the hybrid analog/digital architecture of the full-duplex wideband integrated access and backhaul system.}
    \label{architecture}
\end{figure*}

\subsection{Self-Interference Channel}
The SI channel is decomposed into a static line-of-sight (LOS) channel modeled by $\mathsf{H}_{\mathsf{LOS}}$, which is derived from the geometry of the transceiver, and a non-line-of-sight (NLOS) channel described by $\mathsf{H}_{\mathsf{NLOS}}[\ell]$ which follows the wideband geometric channel model defined by (\ref{channel}). The ($p,q$)-th entry of the LOS SI leakage matrix can be written as 
\begin{equation}\label{eq2.2}
 [\mathsf{H}_{\mathsf{LOS}}]_{pq} = \frac{1}{d_{pq}}\mathsf{e}^{-j2\pi\frac{d_{pq}}{\lambda}}    
\end{equation}
where $d_{pq}$ is the distance between the $p$-th antenna in the TX array and $q$-th antenna in the RX array at BS given by (\ref{distance}). The aggregate SI channel matrix can be obtained by
\begin{equation}\label{eq2.3}
\mathsf{H}_{\mathsf{SI}}[\ell] = \underbrace{\sqrt{\frac{\kappa}{\kappa+1}}\mathsf{H}_{\mathsf{LOS}}}_{\textsf{Near-Field}} + \underbrace{\sqrt{\frac{1}{\kappa+1}}\mathsf{H}_{\mathsf{NLOS}}[\ell]}_{\textsf{Far-Field}}   
\end{equation}
where $\kappa$ is the Rician factor.
\begin{figure*}[t]
\begin{equation}\label{distance}
\begin{split}
d_{pq} = \sqrt{ \left( \frac{d}{\tan(\omega)} + (q-1)\frac{\lambda}{2} \right)^2 + \left( \frac{d}{\sin(\omega)}+(p-1)\frac{\lambda}{2} \right)^2 -2\left( \frac{d}{\tan(\omega)} + (q-1)\frac{\lambda}{2} \right)\left( \frac{d}{\sin(\omega)}+(p-1)\frac{\lambda}{2} \right)\cos(\omega)   }    
\end{split}
\end{equation}
\end{figure*}

\subsection{Quantized Signal Model}
For infinite resolution, a typical received signal is given by
\begin{equation}\label{unqsignal}
    \mathbf{y} = \mathbf{H}\mathbf{x} + \mathbf{n}
\end{equation}
where $\mathbf{H}$, $\mathbf{x}$, and $\mathbf{n}$ are the channel matrix, precoded symbols, and additive white Gaussian noise (AWGN), respectively. Many nonlinear quantization models have been proposed in the literature; however, the analysis of such models is complex for a higher number of ADC bits. In quantized systems, a lower bound on the spectral efficiency has been derived by treating the quantization as additive Gaussian noise with variance inversely proportional to the resolution of the quantizer, that is, $2^{-b}$ times the received input power where $b$ is the number of ADC bits. Recent works \cite{24,28} have considered this additive quantization noise model (AQNM) for mmWave signals with arbitrary numbers of ADC bits. In addition, other works \cite{18,22} derived Gaussian approximations using Bussgang Theory to linearize the nonlinear quantization distortion which is quite similar to AQNM modeling. The received signal (\ref{unqsignal}) is processed through the RF chains and then converted to the digital domain by the ADC. The AQNM represents the quantized version of (\ref{unqsignal}) given by
\begin{equation}\label{quansignal}
\Tilde{\mathbf{y}} = \alpha \mathbf{y} + \mathbf{q}    
\end{equation}
where $\mathbf{q}$ is the additive quantization noise, $\alpha = 1-\eta$, and $\eta$ is the inverse of the signal-to-quantization-plus-noise ratio (SQNR), which is inversely proportional to the square of the resolution of an ADC, i.e., $\eta = \frac{\pi\sqrt{3}}{2} \cdot 2^{-2b}$.  See Table \ref{etavalues}.

\begin{table}[b]
\renewcommand{\arraystretch}{1}
\caption{$\eta$ for different numbers of bits $b$ \cite{22}.}
\label{etaparam}
\centering
\begin{tabular}{cccccc}
$\boldsymbol{b}$ & 1 & 2 & 3 & 4 & 5\\
\hline
$\boldsymbol{\eta}$ & 0.3634 & 0.1175 & 0.03454 & 0.009497 & 0.002499
\end{tabular}
\label{etavalues}
\end{table}

\subsection{All-Digital Beamforming}
Prior to any quantization, the received signals $\mathbf{y}_{\mathsf{backhaul}}[k]$ and $\mathbf{y}_{\mathsf{access}}[k]$ at the $k$-th subcarrier are given by
\begin{equation}
\begin{split}
\mathbf{y}_{\mathsf{backhaul}}[k] =&  \underbrace{\sqrt{\rho_b}\mathbf{W}_{\mathsf{IAB}}^*[k]\mathbf{H}_b[k]\mathbf{F}_{\mathsf{gNB}}[k]\mathbf{s}_b[k]}_{\textsf{Desired Signal}}\\& + \underbrace{\sqrt{\rho_s} \mathbf{W}_{\mathsf{IAB}}^*[k]\mathbf{H}_{\mathsf{SI}}[k]\mathbf{F}_{\mathsf{IAB}}[k]\mathbf{s}_a[k]}_{\textsf{Self-Interference Signal}}\\& +  \underbrace{\mathbf{W}_{\mathsf{IAB}}^*[k] \mathbf{n}_{\mathsf{IAB}}[k]}_{\textsf{AWGN}}
\end{split}
\end{equation}
\begin{equation}
\begin{split}
\mathbf{y}_{\mathsf{access}}[k] =&  \underbrace{\sqrt{\rho_a}\mathbf{W}_{\mathsf{UE}}^*[k]\mathbf{H}_a[k]\mathbf{F}_{\mathsf{IAB}}[k]\mathbf{s}_a[k]}_{\textsf{Desired Signal}} +  \underbrace{\mathbf{W}_{\mathsf{UE}}^*[k] \mathbf{n}_{\mathsf{UE}}[k]}_{\textsf{AWGN}}
\end{split}
\end{equation}
where $\mathbf{W}_{\mathsf{UE}}[k]$, $\mathbf{W}_{\mathsf{IAB}}[k]$, $\mathbf{F}_{\mathsf{gNB}}[k]$, and $\mathbf{F}_{\mathsf{IAB}}[k]$ are the all-digital combiners and precoders at the $k$-th subcarrier for the UE and IAB, respectively.

After the ADCs, the quantized backhaul and access received signals at the $k$-th subcarrier are expressed by
\begin{equation}
\Tilde{\mathbf{y}}_{\mathsf{backhaul}}[k] = \alpha_b \mathbf{y}_{\mathsf{backhaul}}[k] + \underbrace{\mathbf{W}^*_{\mathsf{IAB}}[k] \mathbf{q}_{\mathsf{IAB}}[k]}_{\textsf{AQNM}}  
\end{equation}
\begin{equation}
\Tilde{\mathbf{y}}_{\mathsf{access}}[k] = \alpha_a \mathbf{y}_{\mathsf{access}}[k] + \mathbf{W}^*_{\mathsf{UE}}[k] \mathbf{q}_{\mathsf{UE}}[k]  
\end{equation}
At the $k$-th subcarrier, we define the covariance matrices of the AQNM at the UE and the IAB node, respectively, as follows
\begin{equation}
\mathbf{R}_{\mathbf{q}_{\mathsf{UE}}}[k] = \alpha_a(1-\alpha_a)\mathsf{diag}\left(\rho_a \mathbf{H}_a^{\mathsf{eff}}[k]\mathbf{H}_a^{\mathsf{eff}*}[k] + \mathbf{I} \right)    
\end{equation}
\begin{equation}
\begin{split}
\mathbf{R}_{\mathbf{q}_{\mathsf{IAB}}}[k] =& \alpha_b(1-\alpha_b)\mathsf{diag}\left( \rho_b \mathbf{H}_b^{\mathsf{eff}}[k]\mathbf{H}_b^{\mathsf{eff}*}[k]\right.\\& \left. + \rho_s\alpha_b^2\mathbf{H}_{\mathsf{SI}}^{\mathsf{eff}}[k]\mathbf{H}_a^{\mathsf{eff}}[k]\mathbf{H}_a^{\mathsf{eff}*}[k]\mathbf{H}_{\mathsf{SI}}^{\mathsf{eff}*}[k] +\mathbf{I} \right)   
\end{split}
\end{equation}
where the effective channel is defined by $\mathbf{H}^{\mathsf{eff}} = \mathbf{W}^*\mathbf{H}\mathbf{F}$.
\subsection{Hybrid Beamforming}
For hybrid combining, the backhaul and access quantized received signals after ADC are expressed by
\begin{equation}
\begin{split}
\Tilde{\mathbf{y}}_{\mathsf{backhaul}}[k] =& \alpha_b\left(\sqrt{\rho_b}\mathbf{W}_{\mathsf{IAB}}^{\mathsf{BB}*}[k] \mathbf{W}_{\mathsf{IAB}}^{\mathsf{RF}*}\mathbf{H}_b[k]\mathbf{F}_{\mathsf{gNB}}^{\mathsf{RF}}\mathbf{F}_{\mathsf{gNB}}^{\mathsf{BB}}[k]\mathbf{s}_b[k]\right.\\&\left.+ \sqrt{\rho_s} \mathbf{W}_{\mathsf{IAB}}^{\mathsf{BB}*}[k]\mathbf{W}_{\mathsf{IAB}}^{\mathsf{RF}*}\mathbf{H}_{\mathsf{SI}}[k]\mathbf{F}_{\mathsf{IAB}}^{\mathsf{RF}}\mathbf{F}_{\mathsf{IAB}}^{\mathsf{BB}}[k]\mathbf{s}_a[k]\right.\\&\left.+\mathbf{W}_{\mathsf{IAB}}^{\mathsf{BB}*}[k] \mathbf{W}_{\mathsf{IAB}}^{\mathsf{RF}*} \mathbf{n}_{\mathsf{IAB}}[k] \right)  +  \mathbf{W}_{\mathsf{IAB}}^{\mathsf{BB}*}[k] \mathbf{q}_{\mathsf{IAB}}[k]
\end{split}
\end{equation}
\begin{equation}
\begin{split}
\Tilde{\mathbf{y}}_{\mathsf{access}}[k] =& \alpha_a\left(\sqrt{\rho_a}\mathbf{W}_{\mathsf{UE}}^{\mathsf{BB}*}[k] \mathbf{W}_{\mathsf{UE}}^{\mathsf{RF}*}\mathbf{H}_a[k]\mathbf{F}_{\mathsf{IAB}}^{\mathsf{RF}}\mathbf{F}_{\mathsf{IAB}}^{\mathsf{BB}}[k]\mathbf{s}_a[k]\right.\\&\left.+\mathbf{W}_{\mathsf{UE}}^{\mathsf{BB}*}[k] \mathbf{W}_{\mathsf{UE}}^{\mathsf{RF}*} \mathbf{n}_{\mathsf{UE}}[k] \right)  +  \mathbf{W}_{\mathsf{UE}}^{\mathsf{BB}*}[k] \mathbf{q}_{\mathsf{UE}}[k]
\end{split}
\end{equation}
where $\mathbf{W}_{\mathsf{UE}}^{\mathsf{BB}}[k]$, $\mathbf{W}_{\mathsf{IAB}}^{\mathsf{BB}}[k]$, $\mathbf{F}_{\mathsf{gNB}}^{\mathsf{BB}}[k]$, and $\mathbf{F}_{\mathsf{IAB}}^{\mathsf{BB}}[k]$ are the digital combiners and precoders at the $k$-th subcarrier for the UE and IAB, respectively. 
$\mathbf{W}_{\mathsf{UE}}^{\mathsf{RF}}$, $\mathbf{W}_{\mathsf{IAB}}^{\mathsf{RF}}$, $\mathbf{F}_{\mathsf{gNB}}^{\mathsf{RF}}$, and $\mathbf{F}_{\mathsf{IAB}}^{\mathsf{RF}}$ are the analog combiners and precoders for the UE and IAB, respectively. The hybrid analog/digital beamformers of the system are illustrated by Fig.~\ref{architecture}.

\section{Beamforming Design}
The objective of the beamforming design is to maximize the sum spectral efficiency by canceling the SI and maximizing the power received from the UE. For the all-digital beamformer, the optimization problem can be formulated as follows:
\begin{equation}\label{opt1}
\begin{split}
\mathscr{P}_1:& \max\limits_{\mathbf{F}_{\mathsf{gNB}}[k], \mathbf{F}_{\mathsf{IAB}}[k], \mathbf{W}_{\mathsf{IAB}}[k], \mathbf{W}_{\mathsf{UE}}[k]} {\mathcal{I}_{\mathsf{backhaul}} + \mathcal{I}_{\mathsf{access}}}\\
\mathsf{s.t.}~&\|\mathbf{F}_{\mathsf{gNB}}[k]\|^2_F = N_s^{\mathsf{gNB}},~k=0\ldots K-1 \\&
\|\mathbf{F}_{\mathsf{IAB}}[k]\|^2_F = \|\mathbf{W}_{\mathsf{IAB}}[k]\|^2_F = N_s^{\mathsf{IAB}},~k=0\ldots K-1  \\&
\|\mathbf{W}_{\mathsf{UE}}[k]\|^2_F = N_s^{\mathsf{UE}},~k=0\ldots K-1 
\end{split}
\end{equation}
Here, $\mathcal{I}_{\mathsf{backhaul}}$ and $\mathcal{I}_{\mathsf{access}}$ are the spectral efficiencies of the backhaul and access links, respectively. For hybrid beamforming architecture, the problem (\ref{opt1}) can be reformulated as
\begin{equation}\label{opt2}
\begin{split}
\mathscr{P}_2:& \max\limits_{\substack{\mathbf{F}_{\mathsf{gNB}}^{\mathsf{BB}}[k], \mathbf{F}_{\mathsf{IAB}}^{\mathsf{BB}}[k], \mathbf{W}_{\mathsf{IAB}}^{\mathsf{BB}}[k], \mathbf{W}_{\mathsf{UE}}^{\mathsf{BB}}[k]\\\mathbf{F}_{\mathsf{gNB}}^{\mathsf{RF}}, \mathbf{F}_{\mathsf{IAB}}^{\mathsf{RF}}, \mathbf{W}_{\mathsf{IAB}}^{\mathsf{RF}},\mathbf{W}_{\mathsf{UE}}^{\mathsf{RF}}}} {\mathcal{I}_{\mathsf{backhaul}} + \mathcal{I}_{\mathsf{access}}}\\
\mathsf{s.t.}~&\|\mathbf{F}_{\mathsf{gNB}}^{\mathsf{RF}}\|^2_F = N_{\mathsf{RF}}^{\mathsf{gNB}}\\&
\|\mathbf{F}_{\mathsf{IAB}}^{\mathsf{RF}}\|^2_F = \|\mathbf{W}_{\mathsf{IAB}}^{\mathsf{RF}}\|^2_F = N_{\mathsf{RF}}^{\mathsf{IAB}}\\& \|\mathbf{W}_{\mathsf{UE}}^{\mathsf{RF}}\|^2_F = N_{\mathsf{RF}}^{\mathsf{UE}}\\&
\|\mathbf{F}_{\mathsf{gNB}}^{\mathsf{BB}}[k]\|^2_F = N_s^{\mathsf{gNB}},~k=0\ldots K-1 \\&
\|\mathbf{F}_{\mathsf{IAB}}^{\mathsf{BB}}[k]\|^2_F = \|\mathbf{W}_{\mathsf{IAB}}^{\mathsf{BB}}[k]\|^2_F = N_s^{\mathsf{IAB}},~k=0\ldots K-1  \\&
\|\mathbf{W}_{\mathsf{UE}}^{\mathsf{BB}}[k]\|^2_F = N_s^{\mathsf{UE}},~k=0\ldots K-1 
\end{split}
\end{equation}
where $N_s$ and $N_{\mathsf{RF}}$ are the numbers of spatial streams and RF chains, respectively. The objective of designing the hybrid beamformers is to minimize the distance between the optimal all-digital precoder $\mathbf{F}_{\mathsf{opt}}$ and the product $\mathbf{F}_{\mathsf{RF}}\mathbf{F}_{\mathsf{BB}}$. Equivalently, the precoding (similarly to combining) design problem can be expressed as 
\begin{equation}
\begin{split}
\mathscr{P}_3:&\left(\mathbf{F}^{\mathsf{opt}}_{\mathsf{RF}},\mathbf{F}^{\mathsf{opt}}_{\mathsf{BB}}\right) = \argmin\limits_{\mathbf{F}_{\mathsf{BB}},\mathbf{F}_{\mathsf{RF}}}\|\mathbf{F}_{\mathsf{opt}}-\mathbf{F}_{\mathsf{RF}}\mathbf{F}_{\mathsf{BB}}\|_F\\
\mathsf{s.t.}~&\mathbf{F}_{\mathsf{RF}} \in \mathcal{F}_{\mathsf{RF}}\\&
\|\mathbf{F}_{\mathsf{RF}} \mathbf{F}_{\mathsf{BB}} \|_F^2 = N_s
\end{split}
\end{equation}
which can be summarized as finding the projection of $\mathbf{F}_{\mathsf{opt}}$ onto the set of hybrid precoders of the form $\mathbf{F}_{\mathsf{RF}}\mathbf{F}_{\mathsf{BB}}$ with $\mathbf{F}_{\mathsf{RF}} \in \mathcal{F}_{\mathsf{RF}}$ and $\mathcal{F}_{\mathsf{RF}}$ is the set of feasible RF precoders. Further, this projection is defined with respect to the Frobenius norm. Unfortunately, the complex non-convex feasible set $\mathcal{F}_{\mathsf{RF}}$ makes finding such a projection both analytically (in closed form) and algorithmically intractable. 
To address this shortcoming, we use \textit{Greedy Hybrid Beamforming} (GHB) \cite{greedy} in designing the hybrid analog/digital beamformers.

Once the hybrid beamformers are designed, the SI may not be properly canceled and the residual may render the performance worse. For this reason, we further focus on designing the digital receive filter which is considered the last line of defense to wipe out the SI on each subcarrier.

\begin{lemma}
The digital receive filters at the UE and IAB node that minimize the SI and aggregated thermal and quantization noise, and hence the Mean Square Error (MSE), are Wiener filters or Linear Minimum MSE (LMMSE) receivers $\mathbf{W}_{\mathsf{MMSE}}$. The filter design problem can be defined as 
\begin{equation}
\begin{split}
\mathscr{P}_4:&\mathbf{W}_{\mathsf{MMSE}} = \argmin\limits_{\mathbf{W}}\mathbb{E}\left[\|\mathbf{s}-\Tilde{\mathbf{y}}\|_2^2 \right]
\end{split}
\end{equation}
where $\mathbf{s}$ and $\Tilde{\mathbf{y}}$ are the transmitted and equalized symbols vectors, respectively. 
\end{lemma}
\begin{algorithm}[t]
\caption{Hybrid Beamforming Design}
\label{hybrid-beamforming}
\begin{algorithmic}[1]
\Function{$\mathsf{Pack}$}{$\big[\mathbf{X}\big]_{k=0}^{K-1}$} \funclabel{alg:a} \label{alg:a-line}
    \State $\overline{\mathbf{X}} \gets \big[ \mathbf{X}[0]~\mathbf{X}[1] \cdot\cdot\cdot   \mathbf{X}[K-1]   \big]$
    \State \Return $\overline{\mathbf{X}}$
\EndFunction
\Statex
\State \textbf{Input} $\mathbf{H}_{\mathsf{SI}}[k],\mathbf{H}_a[k],\mathbf{H}_b[k],~k=0,\ldots,K-1$   
\State  $\left[\mathbf{U}_b[k],\mathbf{\Sigma}_b[k],\mathbf{V}_b[k]\right]$ = $\mathsf{SVD}(\mathbf{H}_b[k])$
\State  $\left[\mathbf{U}_a[k],\mathbf{\Sigma}_a[k],\mathbf{V}_a[k]\right]$ = $\mathsf{SVD}(\mathbf{H}_a[k])$
\State $\mathbf{F}_{\mathsf{gNB}}[k] \gets \left[ \mathbf{V}_b[k]\right]_{:,0:N_s^{\mathsf{gNB}}-1}$
\State $\mathbf{W}_{\mathsf{UE}}[k] \gets \left[ \mathbf{U}_a[k]\right]_{:,0:N_s^{\mathsf{UE}}-1}$
\State $\mathbf{W}_{\mathsf{IAB}}[k] \gets \left[\mathbf{U}_b[k]\right]_{:,0:N_s^{\mathsf{IAB}}-1}$
\State $\mathbf{F}_{\mathsf{IAB}}[k] \gets \left[\mathbf{V}_a[k]\right]_{:,0:N_s^{\mathsf{IAB}}-1}$
\State $\overline{\mathbf{F}}_{\mathsf{gNB}} \gets \mathsf{Pack}\left( \left[ \mathbf{F}_{\mathsf{gNB}}[k] \right]_{k=0}^{K-1} \right)$
\State $\overline{\mathbf{W}}_{\mathsf{UE}} \gets \mathsf{Pack}\left( \left[ \mathbf{W}_{\mathsf{UE}}[k] \right]_{k=0}^{K-1} \right)$
\State $\overline{\mathbf{W}}_{\mathsf{IAB}} \gets \mathsf{Pack}\left( \left[ \mathbf{W}_{\mathsf{IAB}}[k] \right]_{k=0}^{K-1} \right)$
\State $\overline{\mathbf{F}}_{\mathsf{IAB}} \gets \mathsf{Pack}\left( \left[ \mathbf{F}_{\mathsf{IAB}}[k] \right]_{k=0}^{K-1} \right)$
\State $\left[\mathbf{F}_{\mathsf{gNB}}^{\mathsf{RF}},\overline{\mathbf{F}}_{\mathsf{gNB}}^{\mathsf{BB}}  \right] \gets \mathsf{Greedy Hybrid Beamforming}\left(\overline{\mathbf{F}}_{\mathsf{gNB}}\right)$
\State $\left[\mathbf{F}_{\mathsf{gNB}}^{\mathsf{BB}}[k]\right]_{k=0}^{K-1} \gets \mathsf{Unpack}\left(\overline{\mathbf{F}}_{\mathsf{gNB}}^{\mathsf{BB}}  \right)$
\State $\left[\mathbf{W}_{\mathsf{UE}}^{\mathsf{RF}},\overline{\mathbf{W}}_{\mathsf{UE}}^{\mathsf{BB}}  \right] \gets \mathsf{Greedy Hybrid Beamforming}\left(\overline{\mathbf{W}}_{\mathsf{UE}}\right)$
\State $\left[\mathbf{W}_{\mathsf{UE}}^{\mathsf{BB}}[k]\right]_{k=0}^{K-1} \gets \mathsf{LMMSE~Receiver}~(\ref{lmmseue})$
\State $\left[\mathbf{W}_{\mathsf{IAB}}^{\mathsf{RF}},\overline{\mathbf{W}}_{\mathsf{IAB}}^{\mathsf{BB}}  \right] \gets \mathsf{Greedy Hybrid Beamforming}\left(\overline{\mathbf{W}}_{\mathsf{IAB}}\right)$
\State $\left[\mathbf{W}_{\mathsf{IAB}}^{\mathsf{BB}}[k]\right]_{k=0}^{K-1} \gets \mathsf{LMMSE~Receiver}~(\ref{lmmseiab})$
\State $\left[\mathbf{F}_{\mathsf{IAB}}^{\mathsf{RF}},\overline{\mathbf{F}}_{\mathsf{IAB}}^{\mathsf{BB}}  \right] \gets \mathsf{Greedy Hybrid Beamforming}\left(\overline{\mathbf{F}}_{\mathsf{IAB}}\right)$
\State $\left[\mathbf{F}_{\mathsf{IAB}}^{\mathsf{BB}}[k]\right]_{k=0}^{K-1} \gets \mathsf{Unpack}\left(\overline{\mathbf{F}}_{\mathsf{IAB}}^{\mathsf{BB}}  \right)$
\State \Return $\mathbf{F}_{\mathsf{gNB}}^{\mathsf{RF}}$, $\mathbf{F}_{\mathsf{IAB}}^{\mathsf{RF}}$, $\mathbf{W}_{\mathsf{IAB}}^{\mathsf{RF}}$, $\mathbf{W}_{\mathsf{UE}}^{\mathsf{RF}}$, $\mathbf{F}_{\mathsf{gNB}}^{\mathsf{BB}}[k]$, $\mathbf{F}_{\mathsf{IAB}}^{\mathsf{BB}}[k]$, $\mathbf{W}_{\mathsf{IAB}}^{\mathsf{BB}}[k]$, $\mathbf{W}_{\mathsf{UE}}^{\mathsf{BB}}[k]$,$~k=0\ldots K-1$
\end{algorithmic}
\end{algorithm}

\begin{figure*}[t]
\begin{equation}\label{lmmseue}
\mathbf{W}_{\mathsf{UE}}^{\mathsf{BB}}[k] = \left[ \frac{1}{\alpha_a}\left( \overline{\mathbf{H}}_a[k]\overline{\mathbf{H}}_a^*[k] + \frac{N_{\mathsf{RF}}^{\mathsf{UE}}}{\mathsf{SNR}_a} \mathbf{I} + \frac{1}{\alpha^2_a\rho_a}\mathbf{R}_{\mathbf{q}_{\mathsf{UE}}}[k] \right)^{-1} \overline{\mathbf{H}}_a[k]  \right]_{:,0:N_s^{\mathsf{UE}}-1} 
\end{equation}
\begin{equation}\label{lmmseiab}
\mathbf{W}_{\mathsf{IAB}}^{\mathsf{BB}}[k] = \left[\frac{1}{\alpha_b}  \left( \overline{\mathbf{H}}_b[k]\overline{\mathbf{H}}^*_b[k] + \frac{1}{\mathsf{SIR}}\overline{\mathbf{H}}_{\mathsf{SI}}[k]\overline{\mathbf{H}}^*_{\mathsf{SI}}[k]
+ \frac{N_{\mathsf{RF}}^{\mathsf{IAB}}}{\mathsf{SNR}_b}\mathbf{I} + \frac{1}{\alpha_b^2\rho_b}\mathbf{R}_{\mathbf{q}_{\mathsf{IAB}}}[k] \right)^{-1} \overline{\mathbf{H}}_b[k]  \right]_{:,0:N_s^{\mathsf{IAB}}-1}
\end{equation}
\vspace*{-.5cm}
\end{figure*}
\begin{theorem}
For low-resolution ADCs, digital LMMSE receivers at UE and IAB nodes are expressed by (\ref{lmmseue}) and (\ref{lmmseiab}) with $\overline{\mathbf{H}} = \mathbf{W}_{\mathsf{RF}}^*\mathbf{H}\mathbf{F}_{\mathsf{RF}}\mathbf{F}_{\mathsf{BB}}$, $\mathsf{SIR} = \frac{\rho_a}{\rho_s}$ and $\mathsf{SNR}_{\mathsf{x}} = \frac{\rho_\mathsf{x}}{\sigma^2}$, $\mathsf{x} \in \{a, b\}$.
\end{theorem}
\begin{proposition}
For infinite-resolution ADCs, we retrieve exactly the dual discrete-time regularized Zero-Forcing precoder derived with respect to the signal-to-leakage-plus-noise-ratio (SLNR) at the IAB node in \cite{ian}.
\end{proposition}
Algorithm \ref{hybrid-beamforming} summarizes the proposed beamforming design.
\begin{remark}
The application of the SVD requires the knowledge of the channel state information (CSI) which is not available in the practice. Due to the limited space, we defer the analysis of the imperfect CSI to a future extension of this work.
\end{remark}
\begin{remark}
For the baseband beamformers, the dimensionality needed for the LMMSE solution is $N_{\mathsf{RF}}^{\mathsf{IAB}} \geq N_s^{\mathsf{IAB}} + N_s^{\mathsf{gNB}}$ to design the combiner at the IAB node to eliminate the SI. This is because $N_s^{\mathsf{gNB}}$ Degrees of Freedom (DoF) are needed for the combiner and another $N_s^{\mathsf{IAB}}$ DoF to cancel the SI.
\end{remark}
\begin{lemma}
For the interference-free infinite-resolution case, the optimal beamformers diagonalize the channel. By applying the SVD successively on all subcarriers, we retrieve the singular values associated with each subcarrier matrix and extract the first $N_s$ modes associated with the spatial streams. The upper bound for backhaul or access links is given by
\begin{equation}\label{upperbound}
\mathcal{I}_{\mathsf{bound}}=  \frac{1}{K}\sum_{k=0}^{K-1}\sum_{\ell=0}^{N_s-1}\log\left(1 + \frac{\mathsf{SNR}}{KN_s} \sigma_{\ell}\left(\mathbf{H}[k] \right)^2  \right)  
\end{equation}
\end{lemma}

\section{Numerical Analysis}
Table \ref{sysparam} gives the parameter values used for the system simulations.  For each case, 5000 channels realizations were generated to perform the Monte Carlo simulation in MATLAB.

\begin{table}[t]
\renewcommand{\arraystretch}{1}
\caption{System Parameters.}
\label{sysparam}
\centering
\begin{tabular}{rl}
\textbf{Parameter} & \textbf{Value}\\
\hline
Bandwidth & 850 MHz\\
Carrier Frequency & 28 GHz\\
Number of Antennas at IAB/gNB & 64 \\
Number of Antennas at UE & 8\\
Number of RF Chains & 4\\
Number of Spacial Streams  & 2\\
SI Power & 40 dB\\
Roll-Off Factor & 1\\
Delay Taps & 20 \\
Cyclic Prefix & 5 (25 $\%$ of Delay Taps)
\end{tabular}
\end{table}
\vspace*{-1cm}
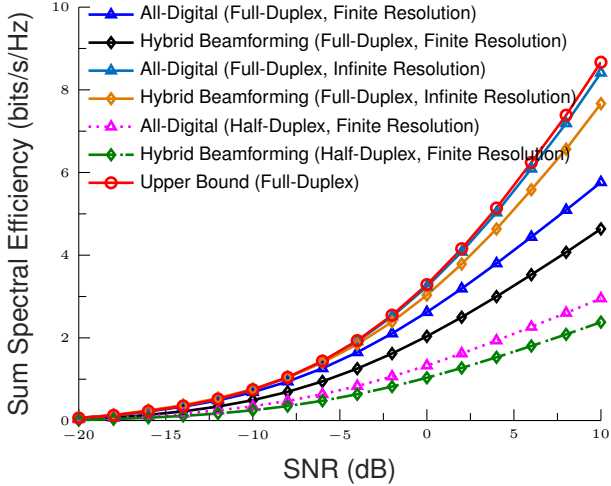
\begin{figure}[t]
\centering
\setlength\fheight{5.5cm}
\setlength\fwidth{7.3cm}
%
%
\definecolor{mycolor1}{rgb}{1.00000,0.00000,1.00000}%
\definecolor{mycolor2}{rgb}{0.00000,0.49804,0.00000}%
\definecolor{mycolor3}{rgb}{0.00000,0.44706,0.74118}%
\definecolor{mycolor4}{rgb}{0.87059,0.49020,0.00000}%
\begin{tikzpicture}

\begin{axis}[%
width=0.951\fwidth,
height=\fheight,
at={(0\fwidth,0\fheight)},
scale only axis,
xmin=-20,
xmax=10,
xlabel style={font=\color{white!15!black}},
xlabel={\sffamily{SNR (dB)}},
ymin=0,
ymax=10,
ylabel style={font=\color{white!15!black}},
ylabel={\sffamily{Sum Spectral Efficiency (bits/s/Hz)}},
axis background/.style={fill=white},
axis x line*=bottom,
axis y line*=left,
legend style={at={(0.0,1.03)}, anchor=north west, legend cell align=left, align=left, draw=none,fill=none}
]
\addplot [color=blue, line width=1pt, mark=triangle, mark options={solid, blue}]
  table[row sep=crcr]{%
-20	0.0610860030862401\\
-18	0.123156262243395\\
-16	0.221136541032006\\
-14	0.334347294052143\\
-12	0.487019080592242\\
-10	0.685665442842903\\
-8	0.941940957067049\\
-6	1.26227460773022\\
-4	1.65009979142279\\
-2	2.10490980530238\\
0	2.62157345570804\\
2	3.19065969883237\\
4	3.80030836670169\\
6	4.43982623020268\\
8	5.09226310066314\\
10	5.76385500415473\\
};
\addlegendentry{\scriptsize\sffamily{All-Digital (Full-Duplex, Finite Resolution)}}

\addplot [color=black, line width=1pt, mark=diamond, mark options={solid, black}]
  table[row sep=crcr]{%
-20	0.0232551558758071\\
-18	0.065767224756509\\
-16	0.137561350670062\\
-14	0.221806470940793\\
-12	0.338260305675877\\
-10	0.49226763274266\\
-8	0.693345377243789\\
-6	0.946743882134663\\
-4	1.25558982138875\\
-2	1.62010777466341\\
0	2.03674762101778\\
2	2.49935754477813\\
4	2.99852617976121\\
6	3.52803677646301\\
8	4.06934751892228\\
10	4.63626343700889\\
};
\addlegendentry{\scriptsize\sffamily{Hybrid Beamforming (Full-Duplex, Finite Resolution)}}

\addplot [color=mycolor3, line width=1pt, mark=triangle, mark options={solid, rotate=0, mycolor3}]
  table[row sep=crcr]{%
-20	0.0614262395291345\\
-18	0.125547510950491\\
-16	0.229178457864902\\
-14	0.35063708165858\\
-12	0.518009908845037\\
-10	0.741643821987569\\
-8	1.03927830341736\\
-6	1.42536895099976\\
-4	1.91345017721111\\
-2	2.51540541096891\\
0	3.23646985606601\\
2	4.07964565731003\\
4	5.02976962458421\\
6	6.09019365861975\\
8	7.18853996795232\\
10	8.41124565973194\\
};
\addlegendentry{\scriptsize\sffamily{All-Digital (Full-Duplex, Infinite Resolution)}}

\addplot [color=mycolor4, line width=1pt, mark=diamond, mark options={solid, mycolor4}]
  table[row sep=crcr]{%
-20	0.0606605802478301\\
-18	0.133529888480543\\
-16	0.244992215307131\\
-14	0.372233906359801\\
-12	0.541420735446206\\
-10	0.759895618013752\\
-8	1.04244516850917\\
-6	1.40032296650999\\
-4	1.84517324638575\\
-2	2.38750919184743\\
0	3.03281219335424\\
2	3.78562889728011\\
4	4.63369599668832\\
6	5.58288785529325\\
8	6.56600229938674\\
10	7.66707972791461\\
};
\addlegendentry{\scriptsize\sffamily{Hybrid Beamforming (Full-Duplex, Infinite Resolution)}}

\addplot [color=mycolor1,dotted, line width=1pt, mark=triangle, mark options={solid, mycolor1}]
  table[row sep=crcr]{%
-20	0.0305603889430557\\
-18	0.0616818680000878\\
-16	0.110881549922326\\
-14	0.167782925751263\\
-12	0.244630634120555\\
-10	0.344790014560724\\
-8	0.474271110464731\\
-6	0.636505901659199\\
-4	0.833463476451316\\
-2	1.06516684491452\\
0	1.32928491205261\\
2	1.62131372682273\\
4	1.93520825604367\\
6	2.2658306011131\\
8	2.60350838948484\\
10	2.95296686739551\\
};
\addlegendentry{\scriptsize\sffamily{All-Digital (Half-Duplex, Finite Resolution)}}

\addplot [color=mycolor2, dash pattern={on 10pt off 2pt on 1pt off 1pt},line width=1pt, mark=diamond, mark options={solid, mycolor2}]
  table[row sep=crcr]{%
-20	0.0117323360086974\\
-18	0.0331256272325327\\
-16	0.0692554893197064\\
-14	0.111678431915497\\
-12	0.170374822950126\\
-10	0.248104513341988\\
-8	0.349780365765038\\
-6	0.478214094147636\\
-4	0.635182197394791\\
-2	0.821049454561761\\
0	1.03425335819219\\
2	1.27193338683722\\
4	1.52928700414273\\
6	1.80346467725559\\
8	2.08406420982319\\
10	2.37957593078769\\
};
\addlegendentry{\scriptsize\sffamily{Hybrid Beamforming (Half-Duplex, Finite Resolution)}}

\addplot [color=red, line width=1pt, mark=o, mark options={solid, red}]
  table[row sep=crcr]{%
-20	0.0614579667875502\\
-18	0.125742746755137\\
-16	0.229791146739349\\
-14	0.351860030903115\\
-12	0.520333327039168\\
-10	0.745854395813471\\
-8	1.04669297916873\\
-6	1.43803693238123\\
-4	1.93441519486503\\
-2	2.54906526461369\\
0	3.28846080570218\\
2	4.1573352804977\\
4	5.14032486816772\\
6	6.24310535544363\\
8	7.38651887500092\\
10	8.66759613640493\\
};
\addlegendentry{\scriptsize\sffamily{Upper Bound (Full-Duplex)}}

\end{axis}
\end{tikzpicture}%
    \caption{Sum spectral efficiency vs. average SNR.  FD methods are shown in solid lines, and HD methods in dashed lines. Finite resolution uses 4 bits.}
    \label{pict1}
\end{figure}

In Fig. \ref{pict1}, the all-digital beamformer outperforms the hybrid beamformer for FD infinite, FD finite, and HD finite resolution cases.  At an SNR of 10 dB, the gaps are about 0.8, 1, and 0.5 bits/s/Hz, respectively. This loss is due to the imperfect GHB projection to design the analog/digital beamformers. When compared to the upper bound at an SNR of 10 dB, the FD finite resolution hybrid beamformer has a gap of about 3.7 bits/s/Hz. The goal of the proposed FD hybrid beamformer is to improve the sum rate over the HD relay, and the FD gain is realized due to the LMMSE digital receiver  at the IAB node to minimize the SI power on each subcarrier.
\begin{figure}[t]
\centering
\setlength\fheight{5.5cm}
\setlength\fwidth{7.3cm}
%
%
\definecolor{mycolor1}{rgb}{0.00000,0.44706,0.74118}%
\definecolor{mycolor2}{rgb}{0.87059,0.49020,0.00000}%
\definecolor{mycolor3}{rgb}{1.00000,0.00000,1.00000}%
\definecolor{mycolor4}{rgb}{0.00000,0.49804,0.00000}%
\definecolor{mycolor5}{rgb}{0.00000,1.00000,1.00000}%
\begin{tikzpicture}

\begin{axis}[%
width=0.951\fwidth,
height=\fheight,
at={(0\fwidth,0\fheight)},
scale only axis,
xmin=1,
xmax=10,
xlabel style={font=\color{white!15!black}},
xlabel={\sffamily{Number of Quantization Bits}},
ymin=0,
ymax=9,
ylabel style={font=\color{white!15!black}},
ylabel={\sffamily{Sum Spectral Efficiency (bits/s/Hz)}},
axis background/.style={fill=white},
axis x line*=bottom,
axis y line*=left,
legend style={at={(0.99,1)}, anchor=south east, legend cell align=left, align=left, draw=none,fill=none}
]
\addplot [color=blue, line width=1.0pt, mark=triangle, mark options={solid, blue}]
  table[row sep=crcr]{%
1	1.6330048470554\\
2	3.16643527040056\\
3	4.59540852447056\\
4	5.8875343702839\\
5	6.96829386471216\\
6	7.73838747069864\\
7	8.24469962182592\\
8	8.51545449851476\\
9	8.66629478982564\\
10	8.6788742641826\\
};
\addlegendentry{\scriptsize\sffamily{All-Digital (Full-Duplex, Finite Resolution)}}

\addplot [color=black, line width=1.0pt, mark=diamond, mark options={solid, black}]
  table[row sep=crcr]{%
1	1.41071223843042\\
2	2.60776324198671\\
3	3.91256203332735\\
4	5.10183943145379\\
5	6.19080421629519\\
6	7.011561510838\\
7	7.59610422851518\\
8	7.93940558543124\\
9	8.15835834040139\\
10	8.19002586341563\\
};
\addlegendentry{\scriptsize\sffamily{Hybrid Beamforming (Full-Duplex, Finite Resolution)}}

\addplot [color=mycolor1, line width=1.0pt, mark=triangle, mark options={solid, rotate=0, mycolor1}]
  table[row sep=crcr]{%
1	8.68990803410697\\
2	8.68036042285194\\
3	8.67572177863935\\
4	8.67218055640588\\
5	8.67252174998593\\
6	8.67855332671921\\
7	8.68328854512074\\
8	8.68517407618065\\
9	8.68906402169664\\
10	8.67982889235709\\
};
\addlegendentry{\scriptsize\sffamily{All-Digital (Full-Duplex, Infinite Resolution)}}

\addplot [color=mycolor2, line width=1.0pt, mark=diamond, mark options={solid, mycolor2}]
  table[row sep=crcr]{%
1	8.17605833492212\\
2	8.16926062371322\\
3	8.16638700530888\\
4	8.1643224080249\\
5	8.16536914041271\\
6	8.17267768749008\\
7	8.17860516583474\\
8	8.18359521913769\\
9	8.19073446462302\\
10	8.19002586341563\\
};
\addlegendentry{\scriptsize\sffamily{Hybrid Beamforming (Full-Duplex, Infinite Resolution)}}

\addplot [color=mycolor3,dotted, line width=1.0pt, mark=triangle, mark options={solid, mycolor3}]
  table[row sep=crcr]{%
1	0.827160418956908\\
2	1.61764684540938\\
3	2.35676624088641\\
4	3.02539706458765\\
5	3.58584180589961\\
6	3.98594810578143\\
7	4.24968223062304\\
8	4.39141055892597\\
9	4.47080935326734\\
10	4.47758955916079\\
};
\addlegendentry{\scriptsize\sffamily{All-Digital (Half-Duplex, Finite Resolution)}}

\addplot [color=mycolor4,dash pattern={on 10pt off 2pt on 1pt off 1pt}, line width=1.0pt, mark=diamond, mark options={solid, mycolor4}]
  table[row sep=crcr]{%
1	0.719198639752801\\
2	1.33698068653607\\
3	2.01373647523875\\
4	2.6307361443997\\
5	3.19679819210034\\
6	3.62382026978709\\
7	3.92786857128108\\
8	4.1063718272641\\
9	4.21975267997938\\
10	4.23505406686377\\
};
\addlegendentry{\scriptsize\sffamily{Hybrid Beamforming (Half-Duplex, Finite Resolution)}}

\addplot [color=cornellred,dotted, line width=1.0pt, mark=triangle, mark options={solid, rotate=0, cornellred}]
  table[row sep=crcr]{%
1	4.48069682342866\\
2	4.47694842598078\\
3	4.47492278362793\\
4	4.47363627648016\\
5	4.47387731644675\\
6	4.47704264132992\\
7	4.4795189161756\\
8	4.48073302117967\\
9	4.4828597136716\\
10	4.47809687587483\\
};
\addlegendentry{\scriptsize\sffamily{All-Digital (Half-Duplex, Infinite Resolution)}}

\addplot [color=green, line width=1.0pt,dash pattern={on 10pt off 2pt on 1pt off 1pt}, mark=diamond, mark options={solid, green}]
  table[row sep=crcr]{%
1	4.22797628751219\\
2	4.22569168505205\\
3	4.22461588953745\\
4	4.22415531514891\\
5	4.22472123180374\\
6	4.22843974156129\\
7	4.23117174641177\\
8	4.2333384838492\\
9	4.23635025629368\\
10	4.23505406686377\\
};
\addlegendentry{\scriptsize\sffamily{Hybrid Beamforming (Half-Duplex, Infinite Resolution)}}

\addplot [color=red, line width=1.0pt, mark=o, mark options={solid, red}]
  table[row sep=crcr]{%
1	8.96139364685732\\
2	8.95389685196157\\
3	8.94984556725586\\
4	8.94727255296033\\
5	8.9477546328935\\
6	8.95408528265985\\
7	8.95903783235119\\
8	8.96146604235934\\
9	8.9657194273432\\
10	8.95619375174966\\
};
\addlegendentry{\scriptsize\sffamily{Upper Bound (Full-Duplex)}}

\end{axis}
\end{tikzpicture}%
    \caption{Sum spectral efficiency vs. the number of quantization bits. The average SNR is set to 10 dB.  FD methods are shown in solid lines, and HD methods in dashed lines.}
    \label{pict2}
\end{figure}
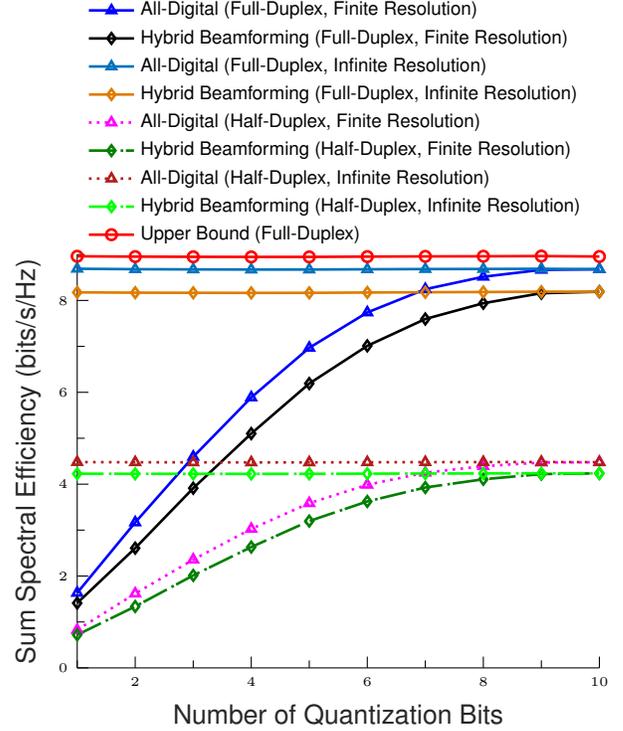

In Fig.~\ref{pict2}, the spectral efficiency degrades at low numbers of quantization bits but converges as the number of bits increases to the ceiling set by the infinite resolution case. The gap between all-digital and hybrid beamformers is less than 1 bit/s/Hz for FD and 0.5 bit/s/Hz for HD over 1-10 bits.  For hybrid beamformers, FD vs. HD gain improves as the number of bits increases; at 4 bits, gain is 2.2 bits/s/Hz. The gap between the FD hybrid beamformer and the upper bound is about 0.8 bit/s/Hz at 9 bits. Our proposed FD design mitigates SI to achieve a significant spectral efficiency gain vs. HD.    

\section{Conclusion}
In this letter, we propose a hybrid beamforming design for wideband IAB based FD systems with low resolution ADCs in a single user scenario to cancel the SI and maximize the sum spectral efficiency. Under low resolution ADCs, the proposed design offers an acceptable FD improvement of around 2.2 bits/s/Hz compared to the HD while the loss due to quantization error when compared to the upper bound is roughly 3.7 bits/s/Hz. For the infinite resolution case, our hybrid beamforming algorithm achieved a small gap in spectral efficiency with the all-digital and upper bound of less than 1 bits/s/Hz indicating the SI is properly suppressed. These results show the feasibility of FD with low resolution ADCs in wideband IAB systems.



\bibliographystyle{IEEEtran}
\bibliography{main}

\end{document}